\documentclass[12pt]{article}

\usepackage{authblk} % authors and affiliations

\usepackage{graphicx} % Required for inserting images
\usepackage[utf8]{inputenc}
\usepackage[total={6.5in, 9in}]{geometry}
\usepackage[skip=10pt plus1pt, indent=0pt]{parskip}

\usepackage{amsmath,amsthm,amssymb,mathtools}

\usepackage{algorithm}
\usepackage{algpseudocode}
\usepackage{xcolor} % http://www.ctan.org/tex-archive/macros/latex/contrib/xcolor
\usepackage{tikz}
\usetikzlibrary{fit,calc}
\newcommand*{\tikzmkleft}[1]{\tikz[remember picture,overlay,] \node (#1) {};\ignorespaces}
\newcommand*{\tikzmkright}[1]{\tikz[remember picture,overlay,] \node (#1) {};\ignorespaces}
\newcommand{\boxitvanilla}[1]{\tikz[remember picture,overlay]{
    \node[  yshift=3pt,
            fill=#1,
            opacity=.25,
            fit={
                ($(Avanilla)-(0.44\linewidth,0.7\baselineskip)$)($(Bvanilla)+(0.75\linewidth,-0.3\baselineskip)$)
            }] {};
    }\ignorespaces
}
\newcommand{\boxitrct}[1]{\tikz[remember picture,overlay]{
    \node[  yshift=3pt,
            xshift=1pt,
            fill=#1,
            opacity=.25,
            fit={
                ($(Arct)-(0.96\linewidth,0.7\baselineskip)$)($(Brct)+(0.68\linewidth,-0.3\baselineskip)$)
            }] {};
    }\ignorespaces
}
\newcommand{\boxitad}[1]{\tikz[remember picture,overlay]{
    \node[  yshift=2pt,
            xshift=1pt,
            fill=#1,
            opacity=.25,
            fit={
                ($(Aad)-(0.71\linewidth,0.7\baselineskip)$)($(Bad)+(0.7\linewidth,-0.13\baselineskip)$)
            }] {};
    }\ignorespaces
}
\newcommand{\boxitrcttwo}[1]{\tikz[remember picture,overlay]{
    \node[  yshift=3pt,
            xshift=10pt,
            fill=#1,
            opacity=.25,
            fit={
                ($(Crct)-(0.82\linewidth,0.7\baselineskip)$)($(Drct)+(0.06\linewidth,-0.14\baselineskip)$)
            }] {};
    }\ignorespaces
}
\newcommand{\boxitrcttwotwo}[1]{\tikz[remember picture,overlay]{
    \node[  yshift=1pt,
            xshift=1pt,
            fill=#1,
            opacity=.25,
            fit={
                ($(Erct)-(0.39\linewidth,0.69\baselineskip)$)($(Frct)+(0.81\linewidth,0)$)
            }] {};
    }\ignorespaces
}
\newcommand{\boxitadtwo}[1]{\tikz[remember picture,overlay]{
    \node[  yshift=3pt,
            xshift=1pt,
            fill=#1,
            opacity=.25,
            fit={
                ($(Aadtwo)-(0.78\linewidth,0.7\baselineskip)$)($(Badtwo)+(0.38\linewidth,-0.24\baselineskip)$)
            }] {};
    }\ignorespaces
}
\newcommand{\boxitadtwotwotwo}[1]{\tikz[remember picture,overlay]{
    \node[  yshift=2pt,
            xshift=1pt,
            fill=#1,
            opacity=.25,
            fit={
                ($(Aadtwotwotwo)-(0.94\linewidth,0.7\baselineskip)$)($(Badtwotwotwo)+(0.83\linewidth,0)$)
            }] {};
    }\ignorespaces
}
\colorlet{pink}{red!40}
\colorlet{blue}{cyan!90}
\colorlet{green}{green!60}

\makeatletter
\renewcommand{\ALG@name}{Method} %Change the name Algorithm to Algoritme
\makeatother

\DeclarePairedDelimiter{\ceil}{\lceil}{\rceil}

\title{Adaptive Experimental Design for\\ Intrusion Data Collection}
\author[1,2]{Kate Highnam}
\author[3]{Zach Hanif}
\author[1]{Ellie Van Vogt}
\author[1]{Sonali Parbhoo} 
\author[1]{ \\ Sergio Maffeis}
\author[4]{Nicholas R. Jennings}

\affil[1]{Imperial College London}
\affil[2]{The Alan Turing Institute}
\affil[3]{Independent Researcher}
\affil[4]{Loughborough University}

\date{\vspace{-2ex}}

\begin{document}
\maketitle
\begin{abstract}
% We apply proven techniques from clinical research to address long-standing issues in cyber intrusion research. 
% Automated cyber intrusion attacks continuously scan and probe internet-connected systems. The state of the art in cyber intrusion defense employs observational techniques augmented with automated statistical techniques, including machine learning (ML). This approach has been effective, but is susceptible to a variety of biases, is limited to specific situations, and is not efficient in its use of resources and time.
% This paper provides theory and empirical data on methods that aim to discover cause-and-effect relationships, and to do so efficiently. The clinical research field provides both inspiration and the theoretical machinery for our work. An adaptive honeypot deployment is executed as a randomized control trial (RCT), with results contrasted with a conventional deployment.

\noindent
Intrusion research frequently collects data on attack techniques currently employed and their potential symptoms.
This includes deploying honeypots, logging events from existing devices, employing a red team for a sample attack campaign, or simulating system activity.
% Historically, intrusion data is collected through observational studies (e.g., by deploying honeypots or logging running devices) or recorded from simulated activity (e.g., from a red team or simulation).
% For example, traditional honeypot deployments expose vulnerable systems for extended periods of time in large quantities.
% Such methods provide data on currently employed attack techniques and their potential corresponding symptoms in a given environment.
However, these observational studies do not clearly discern the cause-and-effect relationships between the design of the environment and the data recorded. Neglecting such relationships increases the chance of drawing biased conclusions due to unconsidered factors, such as spurious correlations between features and errors in measurement or classification. 
In this paper, we present the theory and empirical data on methods that aim to discover such causal relationships efficiently. 
% The clinical research field provides both inspiration and proven theoretical techniques for our work. 
Our \textbf{adaptive design (AD)} is inspired by the clinical trial community: a variant of a randomized control trial (RCT) to measure how a particular ``treatment'' affects a population. 
% While more restrictive than the breadth of questions a traditional deployment could answer, our directed approach quickly answers specific questions such as ``Does inserting this API vulnerability increase the chances of seeing exploits on the SQL server?'' and ``Are attackers constantly exploiting misconfigured cloud instances across cloud regions in the U.S.?''.
To contrast our method with observational studies and RCT, we run the first controlled and adaptive honeypot deployment study, identifying the causal relationship between an \texttt{ssh} vulnerability and the rate of server exploitation. 
% By conducting studies with a control, we uncover (with high confidence) the cause-and-effect relationship a vulnerability has on the likelihood of system exploitation.
% An adaptive design (AD) honeypot deployment is executed as a randomized control trial (RCT), with results contrasted with a conventional deployment.
% % Evidence of causal relationships can be discovered through control-based studies, when a controlled version of an environment is compared to a slightly altered version to determine if the change made caused an event of interest.
We demonstrate that our AD method decreases the total time needed to run the deployment by at least 33\%,  while still confidently stating the impact of our change in the environment. Compared to an analogous honeypot study with a control group, our AD requests 17\% fewer honeypots while collecting 19\% more attack recordings than an analogous honeypot study with a control group.

\end{abstract}

\section{Introduction}

\begin{table*}[t]
    \caption{Mapping of healthcare terminology to security terminology for this paper.}
\begin{center}
\begin{small}
% \begin{sc}
\begin{tabular}{c|c}
    \textbf{Healthcare} & \textbf{Security} \\
    \hline 
    \hline 
    ``trial'' & ``a study comparing honeypots with and without a vulnerability'' \\   
    \hline
    ``study population'' & ``our Ubuntu honeypots with our host-based sensors'' \\ 
    \hline
    ``patient'' or ``participant'' & ``a honeypot'' \\ 
    \hline
    ``recruiting more subjects'' & ``starting more honeypots with specific characteristics'' \\ 
    \hline
    % ``infection'' & ``exploitation'' \\ %``honeypot is exploited'' \\
    % \hline
    ``disease'' & ``attacker technique for exploit'' \\
    \hline
    ``intervention'' or ``treatment'' & ``corruption'' or ``the presence or insertion of a vulnerability'' \\ 
    \hline
    ``treated'' & ``corrupted'' \\
\end{tabular}
% \end{sc}
\end{small}
\end{center}
    \label{tab:terminology}
\end{table*}

Automated cyber intrusion attacks continuously scan and probe internet-connected systems~\cite{spamhaus2022botnets,muir2022realworld}. The state of the art in cyber intrusion defenses employ observational techniques augmented with automated statistical techniques, including temporal point processes and machine learning. 
This approach has been effective, but is susceptible to a variety of biases that might mislead or confuse such solutions from generalizing or learning quicker~\cite{sgaier2020caseforcausalai,howards2018overviewconfounding1}. %, is limited to specific situations~\cite{data centric models, ood papers}, and is not efficient in its use of resources and time~\cite{honeypots}. 
% We can combat these issues by improving the datasets that inform or train such methods.
% These biases have been studied for hundreds of years in healthcare to study the impact of diseases and treatments on patients~\cite{book of why}.
% Using proven techniques from clinical research, we limit the impact of potential bias by improving the datasets that train intrusion detection methods.
We aim to limit the impact of potential bias by improving the datasets that train intrusion detection methods.

% Improving the available datasets can expand their generalisability and limit bias~\cite{}. 
Intrusion datasets can be acquired from third party vendors or compiled by recording logs from existing, simulated, or newly deployed research infrastructure~\cite{darpa98,tavallaee2009nslkdd,shiravi2012iscxids,turcotte2018lanldataset,hamman2020deciphering,andrew2022developing}. 
Conventionally, these methods provide \textbf{observational data}, containing information on current attacks implemented against the given systems and how to observe the attacks in the given environment.
However, even in large volumes, observational data has a high potential for bias due to uncontrolled characteristics, including possible spurious correlations between variables and outcomes or measurement error~\cite{dumitras2011experimental,howards2018overviewconfounding1,howards2018overviewconfounding2,hamman2020deciphering, dhir2021prospective}. 
% Observational data also lacks complete answers to ``What if...?'' and ``Why...?'' questions because there is typically no alternative or counterfactual to compare against within the dataset.
% Answers to such questions would enable statistical models and researchers to understand the impact of changes within the environment
% , e.g., what would change in this malware sample's behavior if we added a new firewall policy to restrict the frequency of successful connections or T

To limit potential erroneous conclusions by both statistical models and researchers, we explore intrusion data collection with a \textbf{control group}: a collection of systems studied that are not altered to compare with identical systems that have been altered. 
% We apply proven techniques from clinical research to address long-standing issues in cyber intrusion research. 
Our usage of control groups in an experimental study draws inspiration from clinical research, one of the oldest fields conducting control-based studies~\cite{bhatt2010evolution,pearl2018book}. 
In healthcare, a typical control group study randomly recruits a subset of a population to remain untreated (as the control) and treated (as the altered version). This is known as a \textbf{randomized-control trial} (RCT), the gold standard for clinical trial methods; its random assignment to groups minimizes the impact of researcher biases while evaluating causal relationships ~\cite{houle2015introduction}. 
Our method is based on \textbf{adaptive design} (AD): a variant of RCT that adds pre-planned opportunities to modify aspects of an ongoing trial in response to data accumulated during the study, without invalidating its integrity~\cite{van2019adaptive, stallard2020efficient,pallmann2018adaptive}. 
RCT and AD both account for known conditions and unforeseen events (e.g., a pandemic or war) which might require the trial to end early by separating the trial into multiple stages to run interim analysis. 

Unlike clinical trials with human patients, intrusion research aims to increase the occurrence of events of interest (i.e., intrusions or exploits). 
See Table~\ref{tab:terminology} for some of the terminology from healthcare mapped to security as it is used to define our work.
To demonstrate our intrusion-focused interventional methods, we use a \textbf{honeypot}, a common tool for recording intrusion data. A honeypot is an intentionally vulnerable system with covert monitoring that is used to both entice and observe attackers without their knowing~\cite{provos2007virtual}. 
% By hosting particular vulnerabilities, honeypots can research modern threats by collecting a high number and wide variety of intrusions. 

Traditional honeypot deployments - or ``vanilla'' deployments as we will call them in this paper - expose a large number of identical vulnerable systems for a particular (extended) length of time to collect intrusion data~\cite{nicomette2011set,alata2006lessons,brew2022threat,provos2007virtual,machmeier2023honeypot,kelly2021comparative}.  % need ch and page for provos...
% For example, some researchers collect attack vectors for over 150 days~\cite{nicomette2011set,alata2006lessons} and deploy over $FIXME$ honeypots~\cite{kelly2021comparative}
% These studies provide a breadth of empirical evidence on current exploitation techniques but at a high cost. 
While sufficiently large and long-lived vanilla deployments all but guarantee observations and can summarize the general state of automated threats, they carry several risks and costs that could be unacceptable. 
If a meaningful quantity of identical honeypots were left online, it would provide an opportunity for adversaries to identify the presence of the employed monitoring tools. 
This can hinder observations (i.e., bias the data) and render the tools useless (i.e., when adversaries stop acting after detecting active monitoring or debugging tools). 
Additionally, large scale deployments cost time and money, which absorbs budget, and can hinder or preclude timely observations.

% \textit{New methods for honeypots must be continuously published to stay ahead of adversarial discovery. Recent works focus on smaller devices in IoT systems...
% % FIXME need more content...

% While sufficiently large and long-lived vanilla deployments all but guarantee observations and can summarize the general state of automated threats, they carry several risks and costs that might be unacceptable at times: 
% \begin{itemize}
% \item Leaving meaningful quantity of identical honeypots online provides opportunity for adversaries to identify the presence of monitoring tools employed within the honeypot. This can hinder observations and render the tools useless, i.e., when adversaries stop acting after detecting active monitoring or debugging tools. 
% \item Honeypots with high fidelity can supply a launchpad to adversaries to further exploit the honeypot to launch other attacks, especially when left running for observing a full attack or a method to disable the monitoring is discovered. 
% \item Tactics, techniques, and procedures employed by attackers rapidly change, necessitating continual monitoring refreshes to ensure that defender knowledge remains current. 
% \item Large scale deployments cost time and money, which absorbs budget, and can hinder or preclude timely observations.
% % Deploying in this manner costs time and money, which can hinder research
% \end{itemize}

In this paper, we present the first control-based deployment method for honeypots to optimize resource allocation and limit honeypot exposure. 
Our method is used in an exemplary study to determine the impact of an \texttt{ssh} vulnerability on cloud servers across the United States.
When compared to the vanilla deployment method with the same setup, we find that AD can determine the impact of the vulnerability in 33\% of the total trial duration, while limiting the likelihood of error. and requesting 17\% fewer honeypots overall. With a control group, our AD collects 19\% more attack recordings than the RCT trial.
% This study is an initial application of a new area of honeypot deployment methodologies.

Our contributions in this work are as follows:

% \begin{enumerate}

\begin{itemize}
    % \item The only data quality heuristics to integrate domain expertise with feature values and relationships to indicate areas for improving the dataset.
    \item The first adaptive method for a control study in security, optimizing resource allocation and duration of the study based on the events seen in prior stages and error tolerance.
    \item The first interventional study using honeypots, demonstrating the effectiveness of our adaptive method and how it helps attribution of environment changes during data collection.
    % \item The only causal analysis of existing open source datasets to identify potentially misleading aspects that could affect ML/posthoc analysis, further motivating our technique for further data collection
\end{itemize}
Although we showcase our method with honeypot deployments, it can be used for other control studies in security applications. For example, one could study the impact of a new spam email training on the rate of spam emails being opened, or on the removal of local file inclusion access on the exploitation rate of a web application hosting other vulnerabilities. Our AD strategy uses a new interpretation of clinical trial methodologies, encouraging infections rather than preventing them mid-trial. Additionally, our study ran using automated scripts, presenting the first fully-automated experimental study. This automation and cheaper application setting enables future inventors of new clinical trial methodologies a new venue to showcase their improvements, rather than run an expensive trial with patients.

This paper is structured as follows. 
Section~\ref{sec:background} reviews control studies in security and provides a brief background on the healthcare-based methods that inspired this work.
We then introduce our new AD method in Section~\ref{sec:adhd} while contrasting it with the vanilla and RCT methods.
In Section~\ref{sec:experiment}, we implement our method against the vanilla and RCT methods in a exemplary honeypot deployment. 
We conclude and consider how the method might not behave the same in other settings in Sections~\ref{sec:conclusions}.
% Section~\ref{study-synopsis} contains a trial synopsis, documenting the experiment run in the style of a healthcare trial, following guidance from the NIH~\cite{nih2017notice}.

%-------------------------------------------------------------------------------
\section{Background} \label{sec:background}
%-------------------------------------------------------------------------------

% This work unifies two major research areas with little historic interaction besides sharing terminology, such as ``infections'' and ``virus.''
% In this section, we begin with the origins of the clinical trial methods that inspired this work and related mappings of these methods to other security domains. We then further justify our choice in study population and review employed honeypot deployment strategies.
% nicomette2011set,machmeier2023honeypot,kelly2021comparative,alata2006lessons
% Traditional honeypot deployments deploy a large number of honeypots for a set amount of time to solely collect empirical data on intrusions. 
% The delay in collection and analysis can also 
% One frequently infers resource constraints as the justification.

% What is a control trial?
An experimental study starts from a hypothesis on how a change or treatment will alter an aspect of a given environment~\cite{peisert2007how}.
The hypothesis is then tested, in the simplest form, by observing a control group that is unchanged and comparing it with another (ideally identical) group that is then changed.
An RCT provides one of the strongest evidence on an intervention's impact due to its random allocation of participants to treatment arms (there might be multiple treatments available) or control arm (standard care or a placebo)~\cite{nhs2021rct,hariton2018randomised}. This process removes potential bias of unaccounted factors in the environment. 
Participants are then observed and their outcomes recorded.

Part of the rigor of RCTs is that all aspects of the trial conduct and (interim) analysis must be documented prior to the execution of the study. This prospective approach avoids the introduction of bias from investigators and statisticians mid-trial. 
An important part of this planning process is the sample size calculation. Using estimates and prior knowledge of interventions, trialists (those conducting the trial) can estimate the required number of participants that must be recruited in order to detect a significant difference in outcomes between the groups~\cite{gupta2016basic}.
% This sample size is redrawn at every stage of the trial; 
After approximating the needs and impacts of the study, the execution of the study should be justified.
% The purpose, required level of human participation, and risks associated with the experiment determines how extensive the prior research and need for the results must be to justify its execution.
% For example, human lives could be saved from an intervention to a disease area, but an experiment is needed to determine the efficacy and applicability of the intervention. 
% If this intervention is found to impact the presence of a disease, i.e. removing symptoms or the disease from the patient.

For medicine, developing the justification to go from drug discovery to licensing can take an average of over 10 years~\cite{vannorman2016drugs}; recent advances in trial methodology have been able to improve the efficiency of trials in order to reduce this time~\cite{wason2019improving}.
The conduct and methodology of a clinical trial are highly regulated because of the direct involvement of patients~\cite{hariton2018randomised}.  However, this is not often a barrier to research in cyber security.
% This arena is both high-stakes and resource-intensive, unlike the cyber security field where infrastructure and regulation are not often barriers to research. 

Security has several advantages in running experimental studies. Digital infrastructure is cheap with the advancement of cloud technologies~\cite{franklin2023digital}. 
Our honeypot study could have cost up to \$2,000USD with 600 participants, compared to the millions of USD required in drug testing~\cite{martin2017how}. 
Digital resources can also be exactly copied as many times as needed, whereas biological studies must make strong assumptions about the similarity between patients. % based on surveys on their background and lifestyles.
% Despite this difference, there are many clear relations between cyber security and 
% we can draw on clinical trial methodologies to improve efficiency in cybersecurity research.
Security experimental studies can be quicker to complete if they consider the attacks that occur at a higher frequency and pace of development than a biological infection or disease.

In this section, we review previous experimental studies in security settings that consider or run control groups. 
We finish this section by briefly highlighting the other techniques developed to improve the classic experimental design that have spawned from the constrained medical setting.

\subsection{Control Trials in Security}

Our work is not the first to apply clinical methodology within security;
prior works focus on the interaction of security and users of digital systems.
For example, Simoiu et al.~\cite{simoiu2019told} survey user awareness of ransomware in the general U.S. population.
Lin et al. \cite{lin2019susceptibility} analyze  spearphishing cyber attacks and its correlation with various human demographics.
A common human-oriented security study involves antivirus software and how it is used by the lay person \cite{somayaji2009evaluating,levesque2018technological,levesque2013clinical}. These works implement an experimental study to find strong indications of how successful antivirus software can be based on human performance. 
Yen et al. \cite{yen2014epidemiological} further extends this research area by incorporating the users job title and responsibilities to contextualize the impact of malware within a company. However, we circumvent the recruitment (and cost) of human involvement by focusing on how these methods can be applied to digital systems with automated, autonomous threats.

Few experimental studies have been published without humans in cyber security.
Bošnjak et al.~\cite{bosnjak2020shoulder} prepare an experimental study to systematically evaluate defenses for shoulder surfing attacks after an extensive literature review.
Gil et al.~\cite{gil2014genetic} approach this by using a case-control study to identify complex relationship of threats to a single host within a large network. Although it is called a ``study,'' a case-control study filters and randomly selects data from purely observational studies for its patient population. There is no interaction with the data collection process.   
Causal relations can be learned from such data but there is no control for error or bias. We discuss how our method controls for error in Section~\ref{sec:adhd}.

These studies indicate a major challenge in security experimental studies: the need for human-interpretable interventions. Recording data in medical settings is relatively straightforward, e.g., heartbeats per minute or body temperature indicating there is a fever. Understanding how these indicate a particular disease is also fairly intuitive. But translating host-based logs to indications of unwanted activity in a system is immensely difficult, let alone stating the type of unwanted activity. Thus, mapping ``symptoms'' from sensor logs can be difficult unless we control our human-level interventions. 
% In all our trials, we are involved with our data collection process because the systems we study require insertion of vulnerabilities and an implementation of automated, detailed monitoring to observe activity.
% Similar to our work, they assume a binary state model to describe attacker activity as the event of interest and evaluate these events using a biology-inspired model. 

\subsection{Advances in Control Trial Methodology}
% Adaptive design and others

Randomization limits the impact of unknown external factors in influencing a participant's chance of receiving a treatment, we therefore expect the baseline characteristics of participants to be similar between studied groups. If there is a concern about some baseline characteristics that may be prognostic, then we can stratify randomization based on these variables without loss of statistical strength~\cite{berger2021roadmap}.

Traditional RCTs are known for their rigor and complete pre-specification of procedure.
A common adaptation to an RCT is the inclusion of stopping rules for efficacy, safety, or futility~\cite{pignon1994early}. If the trial has gathered enough evidence that an intervention is effective, or conversely that the intervention is harmful, then the study can cease, saving resources for a future study or a re-run of the same study with corrections. Similarly, interim checks would also catch if there is not enough evidence of an intervention's effect to reach a significant conclusion~\cite{kumar2016interim}.

Contemporary approaches to running RCTs aim to make the process of evaluating an intervention faster and more efficient~\cite{wason2019improving}.
As mentioned, we implement one such method, adaptive design (AD).
The principle of AD is that it permits certain aspects of a study to be modified intermittently based on available evidence.

Interim data used to inform stopping decisions can also be used to inform an updated sample size calculation, or even to update randomization allocation proportions~\cite{pallmann2018adaptive}. A well-known example of this is the REMAP-CAP study, which has many treatments available across multiple domains for treating community acquired pneumonia, including severe COVID-19, in intensive care unit settings~\cite{angus2020remap}.
Platform trials have also been used to successfully evaluate many different types of interventions simultaneously~\cite{sydes2012flexible}.
Monthly interim analyses are conducted and a Bayesian model is used to update randomization probabilities for new participants entering the trial, so that patients are randomized to treatments that are more likely to benefit them~\cite{ning2010response}.
In the present study, we draw on the concept of response adaptive randomization to optimize the allocation of honeypots.

%-------------------------------------------------------------------------------
\section{Adaptive Design for Security Applications} \label{sec:adhd}
%-------------------------------------------------------------------------------
% Discuss the inputs into Methods~\ref{alg:rct} and \ref{alg:ad}, specifically $\alpha$ and $\beta$ and the proportions of incidence. then go through power analysis and survival analysis as special cases in AD. Exact functions and parameters can go into the appendix

In this section, we define our methods for interventional data collection in security applications.
This setting typically studies adversarial effects, i.e., encouraging intrusions as data are collected. We call the change or treatment (e.g., a drug or surgical procedure) made to a population a \textbf{corruption}\footnote{We chose corruption to remove the benevolent intentions frequently affiliated with ``treatment'' from healthcare settings. A reminder that the translations for other clinical trial terms can be found in Table~\ref{tab:terminology}.}. We shall now enumerate the key terms to document prior to executing a study as we define our AD method.

The \textbf{population} considered in the experiment is assumed to be a device or contained system that is or is not corrupted. 
Deploying a copy of these devices or systems is the same as recruiting a patient into a study.
The goal of the study is to achieve a set of objectives, evident from observing a particular \textbf{event of interest}. One can identify an event of interest through the recorded logs on the population; these events should be clear evidence that the corruption caused some change in system behavior. For example, if the corruption is a new login website and we are interested in its effect on attempted SQL injections, then events of interest should be a record of when an SQL injection occurs.

The \textbf{objectives} of our methods are always two fold:
\begin{enumerate}
    \item Confirm evidence of corruption's impact within the population.
    \item Maximize the recording of  events of interest.
\end{enumerate}  
Returning to the SQL injection example, if we wanted to collect a diverse range of attacks rather than automated repeated uses of the same attack, the events of interest could be only recorded if not seen prior.

Before recording events and running the study, it is crucial to accurately define \textbf{endpoints} to anticipate possible errors or miscalculations.
Similar to clinical trials, we recommend setting an endpoint bounding the trial resources by the given budget.
We also recommend stopping the trial early if the adaptive design tries to deploy a group that is too small. If this occurs in the early stages of the trial, it can indicate the rates of allocation have converged to nothing conclusive. This should be followed with a manual review by human experts.
While it might seem inconsequential, it is good practice to list obvious endpoints. 
This might include recording an unexpected exploit technique or an overwhelming number of exploits that break data collection infrastructure.
% This might include recording unexpected SSH access to the device or successfully escalating to root privileges.
% This endpoint demonstrates the thorough details required in the initial documentation to maintain the robustness of the trial.
All of these details must be defined prior to the study execution to maintain the robustness of the trial.

\begin{figure}[t]
    \centering
\begin{minipage}{0.48\textwidth}
\begin{algorithm}[H]
    \centering
    \caption{Vanilla Observational Study}\label{alg:vanilla}
    \footnotesize
    \begin{algorithmic}[1]
        \Require Budget $b$, Trial Duration $t$ (in hours)
        \Ensure $b, t > 0$
        \State $N = \text{GetNumToDeploy}(b, t)$        
        \State \text{/* Start Trial */}
        \tikzmkleft{Avanilla}
        \State $\text{Deploy}(\text{control}=0,\text{corrupted}=N)$ 
        \State $\text{Wait}(t)$ 
        \State $L = \text{SaveLogs()}$ 
        \State \text{CleanUp()}
        \tikzmkleft{Bvanilla}\boxitvanilla{cyan}%{0.5\linewidth}
    \end{algorithmic}
\end{algorithm}
\begin{algorithm}[H]
    \centering
    \caption{Randomized Control Trial}\label{alg:rct}
    \footnotesize
    \begin{algorithmic}[1]
        \Require Budget $b$, alpha $\alpha$, beta $\beta$, Number of Stages $s$, Stage duration $t$ (hours), proportions of interesting events happening in control $p_1$, proportions of interesting events happening in corrupted $p_2$
        \Ensure $s, b, t > 0$; $0.0 < \alpha, \beta < 1.0$
        \tikzmkleft{Crct}
        \State $N_1, N_2 = \text{PowerAnalysis}(p_1, p_2, \alpha, \beta)$ 
        \For{$i = 0; i < s; i \text{+=} 1$}
        \State \text{/* Start Stage of Trial: $N_1 == N_2$ */}
        \tikzmkleft{Drct}\boxitrcttwo{pink}
        \tikzmkleft{Arct}
        \State $\text{Deploy}(\text{control}=N_1,\text{corrupted}=N_2)$ 
        \State $\text{Wait}(t)$ 
        \State $L = \text{SaveLogs()}$ 
        \State \text{CleanUp()}
        \tikzmkleft{Brct}\boxitrct{cyan}%{0.5\linewidth}
        \tikzmkleft{Erct}
        \State $N_\text{total} += N_1 + N_2$
        \If{\text{isEarlyStop($L, b, N_{\text{total}}$)}} 
            \State $i = s$
        \EndIf
        \EndFor
        \tikzmkleft{Frct}\boxitrcttwotwo{pink}
    \end{algorithmic}
\end{algorithm}
\end{minipage}
\hfill
\begin{minipage}{0.48\textwidth}
\begin{algorithm}[H]
    \centering
    \caption{Adaptive Design Study}\label{alg:ad}
    \footnotesize
    \begin{algorithmic}[1]
        \Require Budget $b$, alpha $\alpha$, beta $\beta$, Number of Stages $s$, Stage duration $t$ (hours), proportions of interesting events happening in control $p_1$, proportions of interesting events happening in corrupted $p_2$
        \Ensure $s, b, t > 0$; $0.0 < \alpha, \beta < 1.0$
        % \State $N_{\text{total}} = \text{PowerAnalysis}(\alpha, \beta)$
        \tikzmkright{Aadtwo}
        \State $N_1, N_2 = \text{PowerAnalysis}(p_1, p_2, \alpha, \beta)$
        \For{$i = 0; i < s; i += 1$}
            \State \text{/* Start Stage of Trial */}
            \tikzmkright{Badtwo}\boxitadtwo{pink}
            \tikzmkright{Aad}
            \State $\text{Deploy}(\text{control} = N_1, \text{corrupted} = N_2)$ 
            \State $\text{Wait}(t)$ 
            \State $L = \text{SaveLogs()}$ 
            \State \text{CleanUp()}  
            \tikzmkright{Bad}\boxitad{cyan}%{0.5\linewidth}
            \State $p_1, p_2 = \text{SurvivalAnalysis}(L)$
            \State $N_1, N_2 = \text{PowerAnalysis}(p_1, p_2, \alpha, \beta)$
            \tikzmkright{Aadtwotwotwo}
            \State $N_\text{total} += N_1 + N_2$
            \If{\text{isEarlyStop($L, b, N_{\text{total}}$)}} 
                \State $i = s$
            \EndIf
        \EndFor
        \tikzmkright{Badtwotwotwo}\boxitadtwotwotwo{pink}
    \end{algorithmic}
\end{algorithm}
\end{minipage}
    % \caption{Caption}
\end{figure}

\subsection{Trial Methodologies}

In this section, we review each method for comparison to our AD before using them in a honeypot study (Section~\ref{sec:experiment}). 
The pseudo code for the traditional observational study (vanilla), RCT, and our AD are presented in Methods~\ref{alg:vanilla}, \ref{alg:rct}, and \ref{alg:ad}, respectively.
See Appendix~\ref{appendix:func-defs} for the definitions of the functions. 
The highlighting indicates the similar lines between the algorithms. 
Notably, the RCT and AD trials are split into $s$ stages with early stopping - shown as the loops on line 2 in both Methods~\ref{alg:rct} and \ref{alg:ad}, highlighted in pink. Each stage deploys some proportion of control and corrupted systems, waits for the stage duration, saves the logs, cleans up the deployment and reviews the logs from the stage to see if an endpoint condition has been reached.
The difference between the standard RCT and our AD is what occurs during the interim update. %$p_1$ and $p_2$

The vanilla deployment (Method~\ref{alg:vanilla}) takes a given budget $b$ and the maximum trial duration\footnote{We say in the methods that this is given in hours, but any time duration is works here.} $t$ to determine the maximum number of devices $N$ that can be observed within this study - noted as \texttt{GetNumToDeploy($b$, $t$)} on line 1. Then $N$ altered devices are then deployed for observation during the ``trial''; no control devices are present.

In contrast, the RCT (Method~\ref{alg:rct}) and AD (Method~\ref{alg:ad}) account for the risk of error into selecting how many devices to study using \textbf{power analysis}\footnote{This calculation can be found in more detail in Appendix~\ref{appendix:func-defs}.}.
We pass four parameters into the power analysis equation from HECT~\cite{thorlund2019highly}: the probability of committing a Type I error ($\alpha$), the probability of committing a Type II error ($\beta$), and the rate of incidence for the control and corrupted groups.
A Type I error means claiming an effect is present due to the corruption when it is not true.
A Type II error means the study did not collect evidence of an effect when it is correct. The \textbf{power} of a study is the inverse of the likelihood of committing a Type II error ($1-\beta$). The rate of incidence for the control and corrupted groups is initially determined by a pilot study or educated guess based on related reports. It is an approximation of the rate an event of interest should be observed within the stage. The power analysis equation returns the total number $N_\text{total}$ of devices needed to deploy in each stage. We equally split this value for RCT and the initial stage of AD. After the first stage of AD, we use the updated rates from the prior stage to weight the split of $N_\text{total}$, adapting the allocation of resources mid-trial.   
% In the medical adaptive design, the KM-provided likelihoods would directly form the new $p_1$ and $p_2$. This would encourage the model to decrease the number of death events seen within the trial, which would make sense in a healthcare setting. However, we wish to increase the number of death events seen so we invert the likelihoods from the KM, which we call the \textbf{risk rates} (RRs). The RRs are marginalised and passed to Equation~\ref{eq:power-analysis} to get our new $N_{total}$. Unlike our initial allocation of honeypots by equally splitting $N_{total}$, the RRs weight the allocation so regions and treatments with higher RR are given more honeypots in the following stage.
% This is repeated at every interim analysis for the duration of the trial.

Based on the responses seen in the previous stage within an AD trial, trialists can use interim analysis to make pre-defined changes that will not invalidate or weaken the power of a study.
Our AD updates the next population counts for control and corrupted.
To not risk weakening the power of our study, we apply this update indirectly between stages through the assumed rates of incidence ($p_1$ and $p_2$).

From the logs we have a complete view into the events of interest for the study. 
% These events are also known as \textbf{death events} in health studies if the goal of the study is to save lives.
Our AD method assumes that each participant will have at most one event of interest before terminating the system, removing it from the trial. In the case of honeypots, this would be to protect the honeypot from becoming a launchpad or providing free resources to attackers.
To calculate the likelihood of an event of interest occurring, i.e., the rate of incidence within a group, we use a \textbf{Kaplan-Meier (KM) Function}, a popular approach for survival analysis within healthcare applications\cite{sullivan,robinson2022two}. 
% The events of interest are parsed out and passed to a survival function which calculates the likelihood of surviving, i.e., for a system to not see the event of interest. 
% We chose to use a \textbf{Kaplan-Meier (KM) Function}, a popular approach for survival analysis \cite{sullivan,robinson2022two}. 
% FIXME get better citations
% This means the new incidence of a honeypot with a particular treatment not being infected.

During a stage, the KM function is updates the likelihood $S$ upon every event of interest recorded. At $t=0$, all participants are at risk and $S(t) = 1$. The remaining time steps update following this: 
\begin{equation} \label{eq:survival-analysis}
    S_{t+1} = S_t \times \frac{N_{t} - D_{t+1}}{N_{t}}
\end{equation}
where $N_{t}$ is the current number of participants at risk and %; equal to the total number of patients not including \\ $C_{t+1}$ (the current number of censored patients) and $D_{t+1}$
 $D_{t+1}$ is the number of participants that have seen events of interest since $t$.
The difference ($N_t - D_{t+1}$) is not always one. If we know exactly when all participants see an event of interest, the KM function is calculated when an infection is recorded. In trials without this ability, a time interval must be set to check with the partipants (e.g., every hour or half-hour) to collect data and see if an event has been recorded.
% If we have full information on when all participants within the study see an event of interest, the KM function is calculated when an infection is recorded.
% In clinical trials without full information on their patients, a time interval is set to check in with the patient (e.g. every month or year) to collect data and see if they are alive.
% These likelihoods are calculated at the $n_{t,r}$ level and then marginalized following Figure~\ref{fig:number-hosts} to acquire our new $p_1$ and $p_2$.

%-------------------------------------------------------------------------------
\section{Adaptive Design for Honeypot Deployments} \label{sec:experiment}
%-------------------------------------------------------------------------------

% use the translations presented in Table~\ref{tab:terminology}
We demonstrate the capabilities of our AD for intrusion data collection in a sample study using honeypots. 
Our method can be applied in other intrusion data collections, but we chose one to illustrate its specific capabilities. 
In this study we analyze the risk of an \texttt{ssh} vulnerability within misconfigured cloud servers. 
This scenario is based on a dataset used as a pilot study for our trials\footnote{This citation is removed for anonymity.}. The dataset contains a variety of attacks via the \texttt{ssh} vulnerability, but we can only empirically infer how the presence of this vulnerability affects its likelihood of exploitation.
Although the presence of an \texttt{ssh} vulnerability is well known to affect the rate of exploitation in a server, this study to emphasizes how our method provides evidence on a corruption's impact and the benefits of its adaptation.

This study includes Methods~\ref{alg:vanilla}, \ref{alg:rct}, and \ref{alg:ad} in separate trials, each attempting to collect evidence of the corruption's impact.
Our budget restricts each trial to a maximum of 200 honeypots (approximately \$650USD) over 12 hours.
As recommended in Section~\ref{sec:adhd}, this threshold is noted as the first of our early stopping criteria. 
Based on the pilot study, we assume an initial rate of incidence in the control to be 0.01 and in the corrupted to be 0.4. 
We ensure the study only considers strong evidence by limiting the chance of error\footnote{It is generally accepted in healthcare settings to set $\alpha=0.05$ and $\text{power}=80\%$ (meaning $\beta=0.2$). We assume a power of 90\% ($\beta = 10\% $) because we know there is a large difference between the control and the corrupted.}, setting $\alpha=0.05$ and $\beta=0.10$.
The remaining details for our honeypot studies are summarized in our Study Synopsis (Section~\ref{sec:study-synopsis}). We then review the results of the study in Section~\ref{sec:results}.

\subsection{Study Synopsis} \label{sec:study-synopsis}

Following the guidance issued for clinical trials~\cite{nih2017notice}, we summarize the characteristics of our study below:
% This study analyzes the risk of an \texttt{ssh} vulnerability when inserted into a server, deployed within a United States cloud provider network. 
% When contrasting these corrupted systems to a control, we compare three deployment methodologies aimed to achieve the following objectives.

\begin{description}
\item[Study Duration]: The maximum total duration per trial is 12 hours. This is applied across the three trial methodologies compared in this study 

% \item[Trial Methodologies for Comparison] :
% \begin{description}
%     \item[Vanilla] : Traditional honeypot deployment strategy of spinning up a large number of only corrupted honeypots and leaving them exposed 
%     \item[Randomized Control Trial (RCT)] : Multiple, identical stages with control comparison
%     \item[Adaptive Design] : Multiple stages, adapt quantities of each sub-population based on the responses seen
% \end{description}   

\item[Objectives]: 
(1) Determine if the corruption causes a significant increase in exploitation rate.
(2) Maximize the exploitation rate for honeypots in the U.S. by region in time specified by the trial or stage.

\item[Endpoints]: 
(a) The maximum number of honeypots that can be recruited into the study is 200. 
(b) The total number of honeypots allocated in the corruption group is below 10 (indicating the event of interest is not recorded frequently enough to study in this duration)
(c) The number of honeypots allocated is identical to the last stage of the AD trial, indicating strong evidence has been collected regarding the current rates of incidence.

\item[Study Population]: Cloud-based honeypots monitored with a kernel-level sensor recording all create, clone, and kill system calls. Each honeypot runs with 1 vCPU, 32GB memory, and Ubuntu 20.04. There are no additional programs or fake activity and no active connection between honeypots. 
They are all hosted by the same large-scale cloud provider within the U.S. that instantiates identical servers with unique, randomly assigned IP addresses upon request.
% They are all hosted by the same large-scale cloud provider that instantiates identical servers  within the U.S. 
The IP address ranges are based on the requested region; our study only considers four regions within the U.S.: \texttt{east-1}, \texttt{east-2}, \texttt{west-1}, and \texttt{west-2}.

\item[Study Corruption and Control]: The corruption is an \texttt{ssh} vulnerability that accepts any password for four fake user accounts mimicking IT support accounts on industrial infrastructure: \texttt{user}, \texttt{administrator}, \texttt{serv}, and \texttt{support}. This corruption is an exaggerated version of a common misconfiguration seen in cloud servers~\cite{muir2022realworld,wagener2011adaptive,abdulsalam2021security}. Control honeypots host the same user accounts but only accept ``password'' as the password. We chose the word ``password'' as the password based on evidence of attackers scanning for it in cloud provider networks
~\cite{kelly2021comparative}.

\item[Event of Interest]: We record an event when a user login is seen in one of the four user accounts. Because we never login or generate fake calls to login, any user login seen is considered malicious. 
% Thus, we state an intrusion has occurred when \texttt{ssh} is seen in the \texttt{task} field and the \texttt{loginuid} is either 1001 (\texttt{user}), 1002 (\texttt{administrator}), 1003 (\texttt{serv}), or 1004 (\texttt{support}). The only other accounts available and seen in our sensor logs are the user account managed by the cloud provider (1000) and the operating system (4294967295).

\item[Measuring Corruption Effect]: This study assumes a binary state model to describe each honeypot as whether an intrusion has or has not occurred. The state of the honeypot is determined by real-time monitoring of the logs to deal with ethical issues that may arise from purposefully exposing compute to adversaries.
An exploit is assumed to not have occurred until this event is seen.
\end{description}

To prevent providing free resources to the attacker or opportunities to launch further attacks, we terminate the instances upon recording an event of interest. 
Although this limits the data we acquire from the study, it satisfies our objectives in recording events of interest. 
This can be re-evaluated in alternative studies based on new objectives. % or if this study had not performed as expected.

Each honeypot functions independently with no communication between the honeypots in the same trial. Their logs are aggregated on a central queuing system within their region (for their trial) and downloaded before termination.
Because we use kernel-level sensors, our implementation can be easily extended for other objectives and vulnerabilities.

%-------------------------------------------------------------------------------
\subsection{Results} \label{sec:results}
%-------------------------------------------------------------------------------

\begin{table}[t]
    \centering
    \caption{Comparison of honeypot deployment methods used in each 12-hour trial.}
    \label{tab:total-counts}
\vskip 0.1in
    \begin{tabular}{c || c | c || c | c}
        \textbf{Method for Trial} & \textbf{Control} & \textbf{Corrupted} & \textbf{Total Deployed} & \textbf{Total Attacks Seen} \\
        \hline \hline 
        Vanilla & 0 & 140 & 140 & \textbf{137} \\ \hline
        RCT & 72 & 72 & 144 & 42 \\ \hline
        AD & 32 & 87 & \textbf{119} & 50 \\
    \end{tabular}
\end{table}

\begin{table}[t]
\caption{The deployed honeypots by region and stage for the RCT and the AD trial. Region names are abbreviated to \texttt{e} for ``east'' and \texttt{w} for ``west''.}
\vskip 0.1in
\label{tab:trial-sizes}
\begin{center}
% \begin{sc}
\begin{tabular}{l r || cccc | cccc | c } % FIXME add number of infected at each stage!
 & &  \multicolumn{4}{c}{ \textbf{Control}} & \multicolumn{4}{c}{\textbf{Corrupted}} & \textbf{Total} \\
 & &  e-1 & e-2 & w-1 & w-2 &  e-1 & e-2 & w-1 & w-2 & \\
\hline
\hline
\textbf{Stage 1}& RCT & 6&6&6&6 & 6&6&6&6 & 48\\
& AD & 6&6&6&6 & 6&6&6&6 & 48\\ \hline
\textbf{Stage 2}& RCT & 6&6&6&6 & 6&6&6&6 & \textbf{48}\\
& AD & 4&4&0&0 & 8&12&8&16 & 52\\ \hline
\textbf{Stage 3}& RCT & 6&6&6&6 & 6&6&6&6 & 48\\
& AD & 0&0&0&0 & 2&5&8&4 & \textbf{19}\\ 
\end{tabular}
% \end{sc}
\end{center}
% \vskip -0.1in
\end{table}

% \begin{figure}[t]
%     \centering
%     \includegraphics[width=\linewidth]{imgs/total.png}
%     \caption{Survival curves for the corrupted instances in all trials. The multi-stage trials are coalesced into a single 4-hour line shown. }
%     \label{fig:survival-curves}
% \end{figure}

\begin{figure}[t]
\centering
\begin{tabular}{cc}
    % trim={<left> <lower> <right> <upper>}
  \includegraphics[width=0.47\linewidth,trim={0 0 0 1.5cm},clip]{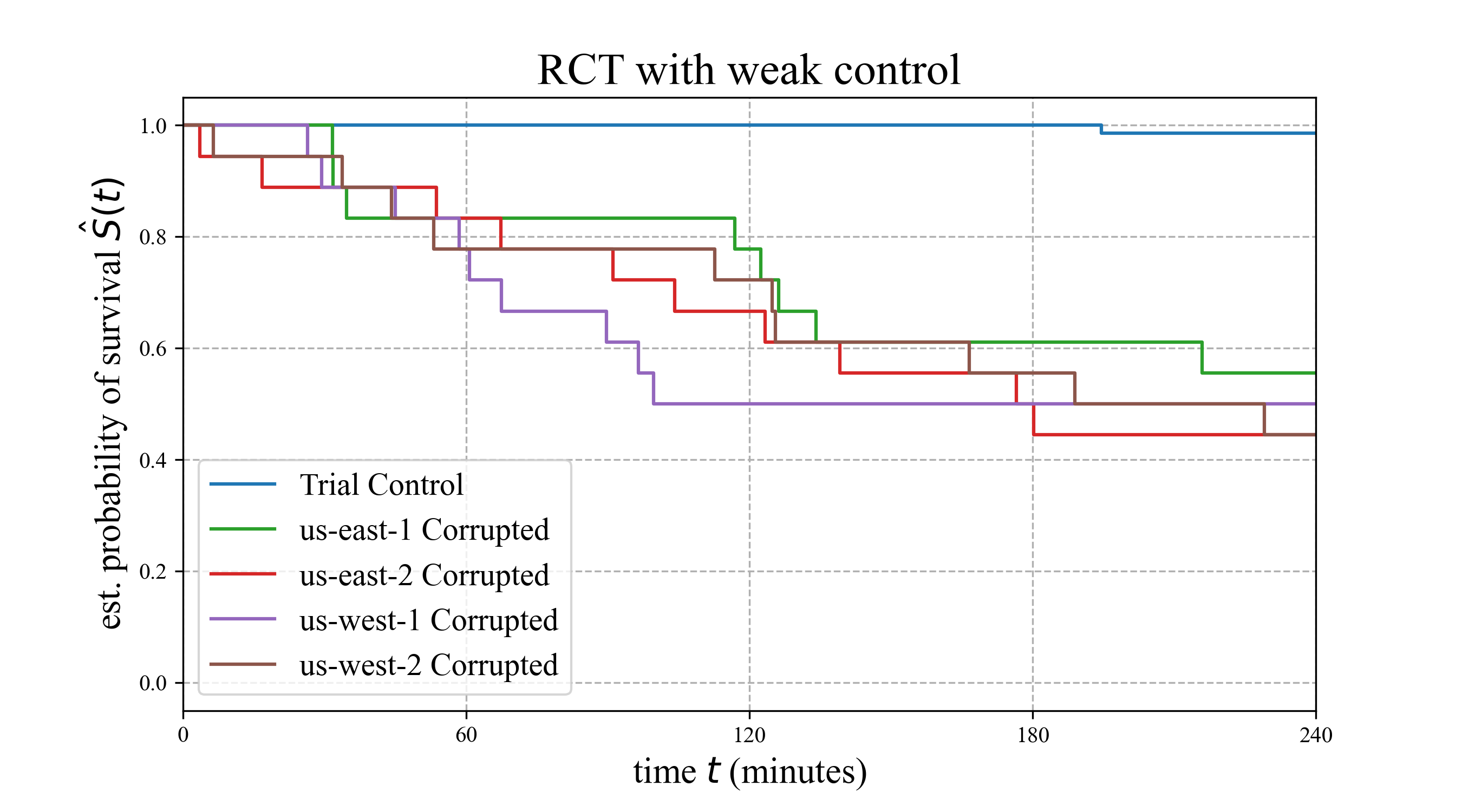} &  \includegraphics[width=0.47\linewidth,trim={0 0 0 1.5cm},clip]{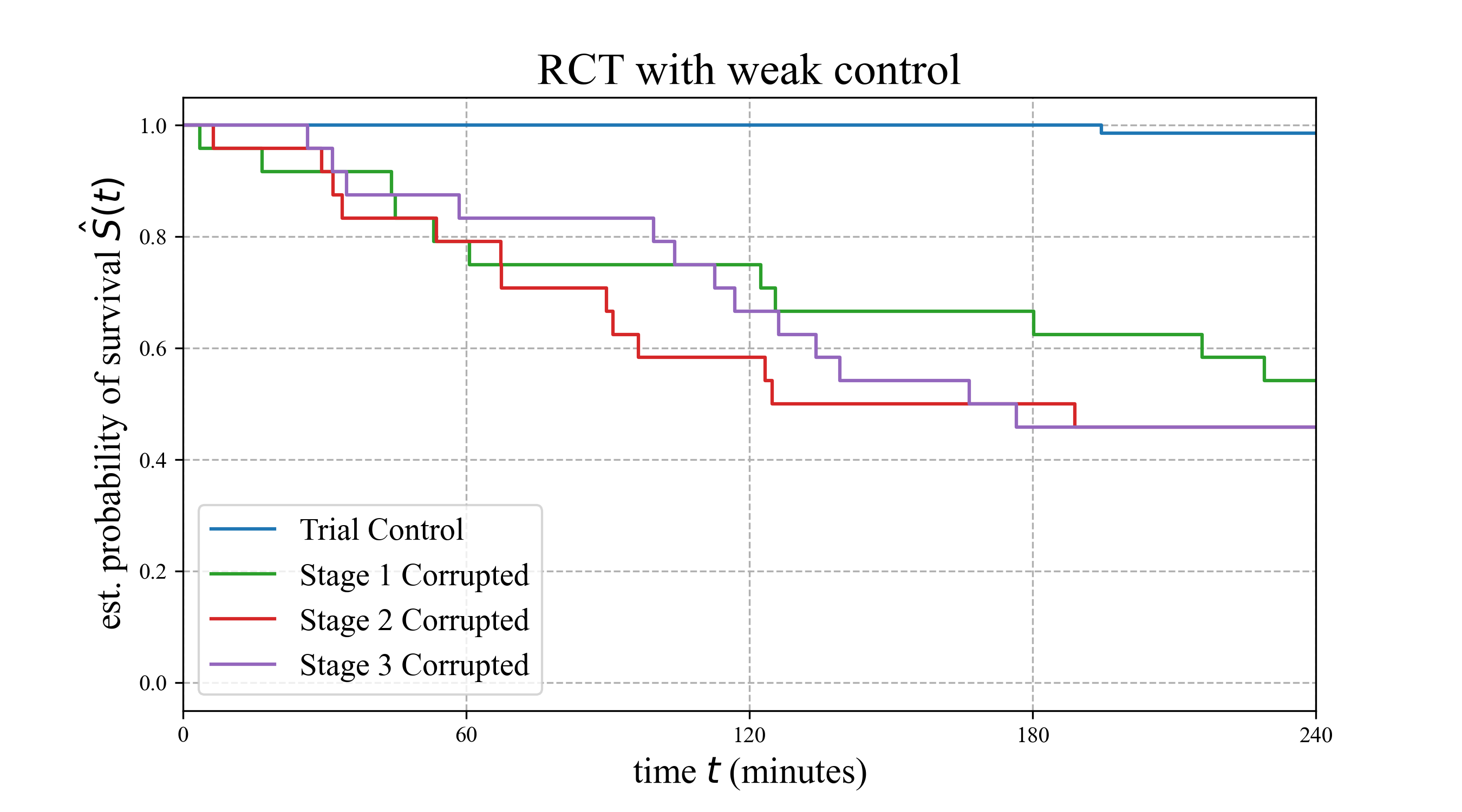} \\
(a) RCT by Region & (b) RCT by Stage \\[6pt]
\end{tabular}
\caption{Comparison of the control and corrupted honeypots by 4-hour stage and region across the RCT trial.}
\label{fig:rct-results}
\end{figure}

% Across all trials and regions in the study, we observed an exponential rate of exploits. 
The total number of honeypots deployed and attacks recorded is shown in Table~\ref{tab:total-counts}. 
As expected, the AD trial deployed the fewest honeypots overall while recording more attacks than RCT.
The AD trial recorded around 36\% of the total attacks seen in the vanilla trial, which saw the highest number of intrusions. This is because the vanilla trial did not deploy any control honeypots, which had a small rate of incidence. However, by not including a control group, it does not account for potential bias in the corruption implementation, preventing it from confidently identifying the causal relationship of the corruption's effect. 
Even so, the data collected could still be used with the rigorous documentation stating the assumptions made in the study. This enables other researchers to review it independently and determine if the study's findings are relevant for their environments. 
% Any bias found later by others can be considered with the results found and knowledge gained in this study is not for naught.

From the trials with a control group, it is clear that the corruption causes an increase in exploitation rate. 
% The logs also show that even though we control for all other characteristics, the exploitation rates were not consistent between trials. 
As can be seen in the trial summary in Table~\ref{tab:trial-sizes}, %this caused the AD trial to request more honeypots for the second stage than RCT.
more control honeypots were exploited than expected in the AD trial, causing a disparity between our initial assumption ($p_1=1\%$) and the results from the first stage (marginalized $p_1=15\%$). 
This caused our AD method to reallocate resources, requesting more honeypots to accommodate for error in the initial assumptions  %according to the observed rate of incidence 
without triggering an endpoint or requiring the study to be reevaluated. 
After the second stage of the AD trial concluded with no exploits in the control group, the control arm was dropped, confirming that the corruption led to more infections.

We could have ended the AD trial after the first stage (saving 66\% of the trial's budget) if Endpoint (b) included the control group in the group size minimum. This was not done because of the second objective to collect intrusion data.
Even with the larger request in stage two, the AD trial requested fewer honeypots than both the vanilla trial and the RCT. 
The stages within the trial also limited the time a honeypot was online, preventing adversaries for extended time to develop a signature for our trap.

Although there was an exponential rate of exploit across the trials, we noticed signs of instability in the four-hour stages, where some regions were observed to have different exploitation rates.
This was especially apparent by comparing their respective survival curves by the regions and stages within the RCT trial, shown in Figure~\ref{fig:rct-results}a and Figure~\ref{fig:rct-results}b. Around 60-120 minutes into the stage, the rate of infection in the regions diverges as though \texttt{us-west-1} was hit first, then \texttt{us-east-2}, \texttt{us-west-2}, and \texttt{us-east-1}, respectively. 
% Although we initially suspected IP scanning to be sequential, there was no obvious pattern in the order of the IP addresses exploited.
From the IP addresses of the hosts, there is no obvious indication of sequential IP scanning.
This instability is another result of this study so future work can note the impact of smaller time windows. 
Because this was unanticipated at the start of this study and not a prior listed early stopping criterion, future work will include it to provide an opportunity for the trialists to discuss whether the trial should continue.

\section{Conclusions} \label{sec:conclusions}
%-------------------------------------------------------------------------------

Our work is the first to apply an adaptive experimental study in intrusion data collection and discuss the benefits of collecting counterfactual information with a control group.
% Control-based studies are frequently used in healthcare to confidently confirm the impact of changes made to a population.
% Our work is the first to apply these proven theoretical methods in intrusion data collection. 
We provide general details on running an experimental study with necessary factors to document prior to conducting the study. Our AD method extends this by optimizing resource allocation based on events seen at every stage, ensuring the statistical confidence through power analysis based on updated exploitation likelihoods with the assumption that an event only occurs once per participant. 
Because the interventional data collected contains true relations between features known through experimentation, future statistical models trained with this data are given higher confidence in learning general trends.
This method is especially applicable for security studies seeking to identify causal relations between a corruption and automated attacks in the wild.
  
% These methods enhance our understanding the effects of changes in a system to cyber risk, but requires some assumptions. 

We then implemented our method in a honeypot study, confirming that the corruption (an \texttt{ssh} vulnerability) increased the infection rate of misconfigured cloud servers. 
% Although we might not be able to say that we have definitively learned about the true infection rate, we confidently fouhave found evidence of variation in the rate by region, as shown by the adaptive design in the weak control trial.
This study also found that while recording more intrusions in observational studies (i.e., in the vanilla trial), the presence of the control group (as in RCT and AD) enables us to identify the corruption effect. Our AD shows it is capable of confirming corruption effect than RCT, requiring only 33\% of the total trial duration to conclude corruption effect and using 17\% fewer honeypots to see 19\% more attacks.
% Although the conclusion that an \texttt{ssh} vulnerability accepting any password increases the rate of exploit compared to only ``password'' is known, 
% this exemplary study shows our AD strategy enables a cheaper and more efficient controlled study.
% Although the vanilla trial observed more exploits, there is no guarantee the corruption caused the observed rate of exploit without a control group. This conclusion is obvious, but this study showcases our deployment strategy. 
Prior to conducting the study, we knew the corruption would increase infection rate because attackers were provided more options for password entry, including the control's ``password'' for the same user accounts.
Had the difference due to the corruption been less apparent (e.g., in altering multiple points of entry or limiting sequences of vulnerability exploits), our study would have taken more time and resources to collect evidence. 
% Deploying control honeypots among the corrupted provides counterfactual information for confidence in the corruption's (causal) effect. 

Future work should consider 
% multiple changes to ensure further consistency between trials and alter the event of interest to consider attack diversity. 
% A simple extension of our work would be to consider multiple cloud providers (like in \cite{kelly2021comparative}) or maximize another event of interest from an \texttt{ssh} vulnerability, such as the collection of malicious artifacts or command and control domain names in network activity.
% We also consider 
implementing multiple vulnerabilities to study the interaction of corruptions. For example, one could add vulnerable  applications within the honeypots to either study the scanning and exploit of multiple existing programs or tracing the sequence of exploits from the \texttt{ssh} vulnerability to a vulnerable application.
This would require introducing a new methodology that can simultaneously consider multiple treatment arms, such as REMAP-CAP~\cite{angus2020remap}. 
% Some primarily focus on treatment interaction, such as REMAP-CAP mentioned in Section~\ref{sec:background}.
By isolating causal relationships, we hope these data can assist in generalizing solutions, remove some bias in the data, and enable other improvements in the intrusion detection community.

%-------------------------------------------------------------------------------
% \bibliographystyle{plain}
\bibliographystyle{unsrt}
\bibliography{main}
%-------------------------------------------------------------------------------

%-------------------------------------------------------------------------------
\appendix
%-------------------------------------------------------------------------------
%-------------------------------------------------------------------------------

\section{Function Definitions} \label{appendix:func-defs}

Pseudo code function definitions used in Methods~\ref{alg:vanilla}, \ref{alg:rct}, and \ref{alg:ad}.

Shared functions:

\texttt{Deploy(control=$n_1$, corrupted=$n_2$)}: Given the number of devices to observe in each group of the study ($n_1$ control devices and $n_2$ corrupted devices), deploy them and record logs from all devices in a centralized location. 

\texttt{Wait($t$)}: Wait the length of time specified by $t$.

\texttt{L = SaveLogs()}: Fetch logs and return them to be stored in variable $L$

\texttt{CleanUp()}: Shut down devices and clean up any infrastructure not kept at the end of the trial or stage.

\texttt{GetNumToDeploy($b$, $t$)}: Given budget $b$ and the maximum trial duration $t$, it returns the maximum number of devices $N$ that can be observed within this study.

\texttt{isEarlyStop(L)}: Given the logs collected from the last stage, determine if the pre-specified early stopping conditions have been met and the trial must immediately terminate.

\texttt{PowerAnalysis($p_1, p_2, \alpha, \beta$)}: See Section~\ref{sec:sample-size}

\texttt{SurvivalAnalysis(L)}: See Section~\ref{sec:survival-analysis}

\subsection{Power Analysis for Sample Size} \label{sec:sample-size}

We denote the trial as robust when it follows a strict calculation of the sample size to be deployed accounts for type I and type II error in the data. For this paper, we follow the equation from HECT~\cite{thorlund2019highly} which follows \textbf{power analysis}  to calculate the sample size:

\begin{equation} \label{eq:power-analysis}
    N_{\textrm{total}} = 2 * \ceil[\Bigg]{\frac{((p_1 q_1 + p_2q_2)(Z_{1 - \alpha/2} + Z_{1-\beta})^2}{(p_1 - p_2)^2} }
\end{equation}

\begin{description}
    \item[$N_{\textrm{total}}$] = Total sample size for the study group, which is later split across the different treatments and regions. We alter the equation to calculate two arms of studies as the same since the split will depend on the ratio found during the trial in interim analysis.
    \item[$Z$] = critical $Z$ value for a given $\alpha$ and $\beta$-based subscript
    \item[$p_1$] = Control Incidence: The assumed rate of the outcome occurring in the control group is initially $p_1=1\%$. 
% FIXME is there a reference to back this choice up? Or is it standard practice for this type of work?
    \item[$p_2$] = Treatment Incidence: The assumed rate of an outcome occurring in the corrupted group based on our pilot study,\footnote{Citation removed for anonymity.} and conservative rounding is $p_2=40\%$. 
    \item[$q_1$, $q_2$] = $1 - p_1$, $1 - p_2$ (respectively)
    \item[$\alpha$] = Type I Error: The probability of claiming an infection rate when it is not true. Our setting and the generally accepted probability in clinical studies is $\alpha= 5\%$.
    \item[$\beta$] = Type II Error: The probability of not detecting an accurate infection rate when it is correct. The inverse ($1 - \beta $) is known as the \textbf{power} of the study. In this study,\footnote{Generally accepted in clinical studies as $power = 80\%$ which means $ \beta = 20\% $.} we assume a power of 90\% ($\beta = 10\% $) because we know there is a large difference between the control and the corrupted. 
% KH: @EV just cuz? small size? - power is typically chosen to be 80-90 % so i wouldn't say this is unusualy high, it's the higher end of standard practice. But also because we know there is a big difference between treated and controls, so if we don't choose a high power the sample size will be tiny, but i don't think you need to explain this
\end{description}
    
In the medical adaptive design, the KM-provided likelihoods would directly form the new $p_1$ and $p_2$. This would encourage the model to decrease the number of death events seen within the trial, which would make sense in a healthcare setting. However, we wish to increase the number of death events seen so we invert the likelihoods from the KM, which we call the \textbf{risk rates} (RRs). The RRs are marginalised and passed to Equation~\ref{eq:power-analysis} to get our new $N_{total}$. Unlike the RCT allocation of honeypots (equally splitting $N_{total}$ between control and corrupted), AD uses the RRs to weight the allocation so regions and corruption assignment with higher RR are given more honeypots in the following stage.
This is repeated at every interim analysis for the duration of the trial.

\subsection{Survival Analysis for Updating Risk Rates} \label{sec:survival-analysis}

We assume in our AD study that each participant will have at most one event of interest before terminating the system, protecting it from becoming a launchpad or providing free resources to attackers.
The events of interest are parsed out and passed to a survival function which calculates the likelihood of surviving, i.e., for a system to not see the event of interest. 
We chose to use a \textbf{Kaplan-Meier (KM) Function}, a popular approach for survival analysis \cite{sullivan,robinson2022two}. 
% FIXME get better citations
% This means the new incidence of a honeypot with a particular treatment not being infected.

The KM function is calculated by updating the likelihood $S$ upon every event of interest recorded. At $t=0$, all participants are at risk and $S(t) = 1$. The remaining time steps update following this: 
\begin{equation}
    S_{t+1} = S_t \times \frac{N_{t} - D_{t+1}}{N_{t}}
\end{equation}
where
\begin{description}
    \item  $N_{t+1}$ is the current number of participants at risk %; equal to the total number of patients not including \\ $C_{t+1}$ (the current number of censored patients) and $D_{t+1}$
    \item  $D_{t+1}$ is the number of participants that have seen events of interest since $t$
\end{description}

\noindent
If we have full information on when all participants within the study see an event of interest, the KM function is calculated when an infection is recorded.
In clinical trials without full information on their patients, a time interval is set to check in with the patient (e.g. every month or year) to collect data and see if they are alive.

\end{document}